# HIDDEN FERMI LIQUID: THE SECRET OF HIGH Tc CUPRATES

**Astract—I present a formalism for dealing directly with the effects of the Gutzwiller projection, which is implicit in the t-J model widely believed to underly the phenomenology of the high-Tc cuprates. I suggest that a true BCS condensation from a Fermi liquid state takes place, but in the unphysical space prior to projection. The theory which results upon projection does not follow conventional rules of diagram theory and in fact in the normal state is a Z=0 non-Fermi liquid. Anomalous properties of the 'strange metal" normal state are predicted and compared to experiments.**

One of the most striking dilemmas about the high Tc superconducting cuprates is that they seem to obey the standard BCS quasiparticle theory when they are superconducting---after all is said and done and in a restricted sense---but that the normal state is undoubtedly not the Fermi liquid on which BCS theory is based.

What I propose here is that there is actually an underlying "hidden Fermi liquid" which undergoes a BCS pairing transition; but that this Fermi liquid is not the usual perturbative continuation from a sea of free electrons. In particular, in the normal state, creating a free, bare electron of momentum k has zero amplitude for creating the hidden Fermi liquid's quasiparticle—that is, the wave function renormalization constant $Z = 0$. For purposes of clarity let us introduce a term for these "hidden" excitations: "pseudoparticles". Many-pseudoparticle excitations saturate the electron's amplitude.

It is then obvious why the hidden Fermi liquid emerges when the substance becomes superconducting: that opens up a gap $\Delta$ for single-pseudoparticle excitations, so that the many-pseudoparticle continuum can only begin at three times that gap. Therefore the total amplitude for many-particle excitations remains finite, and there is a finite Z, although it does not obey the usual perturbation-theory identities relating it to self-energy.

All this is a consequence of the fact that the low-lying electron states of a doped Mott insulator are *projective*, in the sense that the strong repulsion U forces a finite density of the many-electron states to split out of the band continuum of states to the high-energy side, forming the "upper Hubbard band". The remaining continuum is overcompletely described by a perturbation theory based on the conventional theory of a tight-binding band, and which can undergo conventional BCS pairing. The quasiparticle excitations of this "hidden Fermi liquid" are excitations of the physical system, but they are not true quasiparticles of that system because the true particle operators operate on the projected system, not the overcomplete wave function prior to projection. As a result the true operators create multi-particle states of the preprojection system.

FORMALISM
There is much evidence—such as the presence of an actual Mott insulator in the phase diagram—that the interactions in the cuprates are dominated by an on-site Mott-Hubbard term U. Clearly we must deal with U before bothering with relatively minor terms, so we leave other interactions to be accounted for later.

Di Castro et al[1] have demonstrated that any interacting Fermion system which is not a band insulator, and in which none of the interactions is larger than an upper cutoff energy (such as the bandwidth, for tight-binding models), may be renormalized by Shankar's "poor-man's renormalization" scheme into a Fermi-liquid based theory where the relevant Fermionic excitations all can be based on quasiparticles inhabiting a shell around a Fermi surface in momentum space. Quite clearly, the Mott insulator violates this very general theorem; in many more subtle ways, the physics of the doped cuprates in the region of the superconducting dome violates the "poor-man's" prescription. We believe that the reason is that this is a case in which the on-site Coulomb interaction, the Mott-Hubbard interaction U, is larger than the bandwidth and cannot be renormalized to zero, as in Shankar's treatment of Fermi liquid theory.

T M Rice and his co-workers [2] pioneered the correct way to deal with the interaction U when it cannot be renormalized to zero. This is to renormalize it perturbatively to ∞ by means of a canonical transformation exp(iS) which may be determined perturbatively in orders of the kinetic energy hopping matrix element t divided by U. That is, we start with the Hubbard Hamiltonian

$$H = H_0 + H' + H''$$

$$H_0 = U \sum n_{i\uparrow} n_{i\downarrow}; \quad H' = P \sum t_{ij} c^*_{i\sigma} c_{j\sigma} P$$

$$H'' = T - H'$$

[1]

and eliminate successive orders in H" by using the expansion

$$H_{eff} = e^{iS} H e^{-iS} = H + i[S, H] + \cdots \quad [2]$$

and requiring that $H_{eff}$ have no matrix elements to or from the doubly-occupied states.

The result is, with various distracting terms dropped out, the t-J Hamiltonian

$$H_{t-J} = P \sum_{i<j,\sigma} t_{ij} c_{i,\sigma}^* c_{j,\sigma} P + \sum_{i,j} J_{ij} S_i \cdot S_j \quad [3]$$

Here P is the full Gutzwiller projector,

$$P = \prod_i (1 - n_{i,\downarrow} n_{i,\uparrow}) \quad [4]$$

which eliminates all doubly-occupied states. There is no need to apply it to the J term because that does not change site occupancies.

The "distracting" terms which are ignored are, [1], correlated hopping terms caused by the transformation which modify the $t_{ij}$ somewhat (Edegger[3] has estimated their effect in some detail); [2] phonon-mediated interactions, which to lowest order simply modify U, the remainder being small compared to J; [3] the longer-range coulomb interactions which play a role in the mesoscopic charge instabilities (like stripes) which are manifest for low doping and low temperatures, but don't affect the big picture. All can be included as extra terms in the projective Hamiltonian [3] (but must be projected).

Low-energy solutions of [3] are necessarily of the form

$$\Psi = P\Phi(r_1, r_2, \cdots, r_N), \text{ where } \Phi \text{ is a general function of all electron positions.} \quad [5]$$

The Schrodinger equation for $\Psi$, $H_{t-J}\Psi = E\Psi$, may be considered either as an equation for $\Psi$; or, alternatively, as an underdetermined equation for $\Phi$, $H_{t-J}\Phi = E\Phi$ (since E and P of course commute—or, alternatively, we can simply not apply the final projection). We take the point of view that we are attempting to find the $\Phi$ which gives us the best variational energy.

Of course, since the equations are underdetermined, we must add some additional constraint on $\Phi$. What we choose to do is to assume a Hartree-Fock-BCS form for our trial wave function:

$$\Phi = \prod_k (u_k c_{-k,-} + v_k c^*_{k,+})(u_k c_{k,+} + v_k c^*_{-k,-}) | vac >$$

$$= \prod_k \gamma_k \gamma_{-k} | vac >$$

$$= \prod_k (u_k + v_k c^*_{k,+} c^*_{-k,-}) | vac > = \Phi_{BCS} \qquad [6]$$

We define the γ's by the usual Hartree-Fock procedure of requiring that all quasiparticle energies be positive, which gives us a set of 'gap" equations,

$$\langle [H, \gamma^*_k] \rangle \Phi_0 = E_k \gamma^*_k \Phi_0 \qquad [7]$$

The "average" brackets mean that in the three-Fermion terms coming from commutators with the exchange interaction terms we keep all possible mean field terms, i e both the Fock and the BCS self-energies, which we evaluate self-consistently. These are the same equations as given in the 'plain vanilla" paper[4] and its earlier sources.

The Hartree-Fock procedure gives us a unique answer: because the Hamiltonian is translationally invariant we can classify the "pseudoparticles" with a momentum variable and they are orthogonal (I emphasize that they are not true quasiparticles of the full system). In the absence of a gap they will have a Fermi surface obeying Luttinger's theorem. (projection commutes with particle number.) This system of excitations is, in fact, the "hidden Fermi liquid."

Human ingenuity has not provided us with any more flexible trial functions than the Hartree-Fock-BCS product function, and in fact in many-body theory so far we have found this to be quite useful, at least for a start for perturbation theory. You might suggest that Jastrow functions, as in the Laughlin theory, might be used, but in fact that is exactly what we are doing with the canonical transform and the projection: we *are* augmenting our product function with a Jastrow-like correction.

But excitations thus created, although they are valid *excitations* of the projected Hamiltonian, are not *quasiparticles* of the projected problem. They are valid excitations because the Hamiltonian which applies to them is projected and components of the wave function in the forbidden subspace are irrelevant. These excitations of the projective Hamiltonian may be described as $Pc^*\Phi_0$ or $Pc\Phi_0$, but they are quite different from the excitations which are created by applying a particle field operator to the projected state, $c^*P\Phi_0$ or $cP\Phi_0$, and in fact are not created with finite amplitude in this way. This is the essence of the "Strange Metal" phenomenon and will be discussed in detail below: there is a "hidden Fermi liquid" but it is not accessible with one- or two-particle perturbations such as particle tunneling or electromagnetic field.

Counterintuitively, opening a BCS gap makes the initial assumptions better, not worse. The failure of standard Landau theory in the Fermi sea case is caused by the fact that the spectrum of excitations with momentum perpendicular to the Fermi surface has a high density at zero energy, and allows the emission of divergent numbers of low-energy hole-electron pairs. As Haldane has shown,[5] these may be reorganized into Tomonaga bosons for each patch of the Fermi surface, representing the fluctuations of the Fermi surface itself. The density of such pair states is reduced to zero if there is a gap $\Delta$, or equivalently the low-energy Tomonaga bosons disappear. The anomalous quasiparticle decay is replaced by a broad structure around $3\Delta$ or higher, seen in ARPES and tunneling data, that is always referred to as the "hump" and given many bizarre interpretations. Thus we are proposing that there are genuine quasiparticles at low energies in the superconducting state at absolute zero. They have a finite renormalization constant Z but its value does not have the expected relationship to the effective mass or Fermi velocity via the self-energy.

There is an aspect of self-consistency here: when the gap opens the system can exhibit true quasiparticles without anomalous scattering, and this in turn means that the gap may open more easily—i e the superconducting phase gains a little stability by destroying some of the anomalous scattering of the "strange metal" phase. This is a faint residue of the mechanism which was the central idea in my 1997 book[6]. Here we will postpone it until the theory is in a much more quantitative shape.

I have shown[7] how to calculate the Green's functions of the "normal" metal phase of the cuprates, and compared the result to new, accurate experimental data. (in principle the same techniques can be used to describe the superconductor, but the relevant calculations have not yet been done). The methods used are closely related to those for x-ray spectroscopy in metals, which in turn depend on two basic theorems about responses of Fermi systems to scattering potentials, the very well-known Friedel theorem, and the somewhat more obscure Nozieres theorem[8]. These are, first:

$$\delta N = \sum_c \eta_c / \pi \quad [8]$$

Here $\delta N$ is the change in the number of particles surrounding a scattering perturbation which causes phase shifts $\eta_c$ in the scattering channels c. The only channels which affect N are those at the Fermi surface (if there is one). If there is a gap, the scatterer can only change N by creating a bound state, which is equivalent to a phase shift of π, and the change in N is a simple matter of asking how many bound states have been created out of the given band. I have deliberately expressed this theorem in "channel" language because in this way its generality is clear—its obvious relationship to Luttinger's theorem, for instance.

The second theorem is:

$$O(t) \propto t^{-p}, \text{where} \quad p = \sum_c \frac{1}{8} \frac{\eta_c^2}{\pi^2} \quad [9]$$

Here O(t) is a kind of "vacuum" Green's function defined as the overlap integral between the state just prior to the introduction of the scattering phase shifts $\eta_c$ at t=0, and the state at time t later:

$$O(t) = \langle e^{iH_0 t}\Psi_0, e^{i(H_0+V)t}\Psi_0 \rangle, \text{ where } \Psi_0 \text{ is the ground state of } H_0 \text{ and V is the potential which produces the phase shifts } \eta_c. \quad [10]$$

Equations [8] and [9] are what I have called "node-counting theorems" and are unique to Fermi surface systems. In such systems it is not correct to think in terms of fixed numbers of particles because when particles at the Fermi surface are scattered they can move through the boundary. [8] tells us what fractional number of particles is moved through the boundary by a given scattering potential; and [9]—or its equivalent "orthogonality catastrophe" version in terms of sample volume—tells us the penalty in terms of wave function incoherence we must pay for doing so.

That this overlap integral is the relevant correction to the free-quasiparticle Green's function is the central element of the

"strange metal" theory of reference [6], and the derivation is worth repeating in more detail. So far, this has only been applied to the case where the system is normal, there is no anomalous self-energy, and the quasiparticle operators are simple Fermions c and c*, but the generalization to Bogoliubov quasiparticles is direct if complicated: therefore this is the only case we will consider here.

We designate the "real" Fermions which represent physical creation and destruction operators acting in the projected subspace by "hat" operators which do not create or destroy any doubly-occupied sites. These are easily seen to be

$$\hat{c}_{i\sigma} = (1 - n_{i-\sigma})c_{i\sigma} \quad (\hat{c}^*_{i\sigma} = c^*_{i\sigma}(1 - n_{i-\sigma})) \quad [11]$$

Here the hatless operators are to be thought of as operating within the unprojected space, that is they operate on the hidden Fermi liquid. The Green's function for inserting the hole at time 0 and removing it at time t, then, might be written as

$$G_{ii}(0,t) = \langle 0 | \hat{c}^*_{i,\sigma}(0) \hat{c}_{i,\sigma}(t) | 0 \rangle$$
$$\cong \langle 0 | c^*_{i,\sigma}(t) c_{i,\sigma}(0) | 0 \rangle G_{-\sigma}^*(t)$$
$$= G_0(t) G^*(t) \quad \text{where} \quad [12]$$
$$G^* = \langle 0 | (1 - n_{i,-\sigma}[t])(1 - n_{i,-\sigma}[0]) | 0 \rangle$$

here $G_0$ is the free-Fermion Green's function, which for the single-site case of [12] is proportional simply to 1/t. (There are normalization corrections, different actually for electrons and holes, which may be calculated using fugacity factors[9] but are irrelevant to lineshapes.) What we now want to show is that G*, the density-density Green's function, is a Nozieres-de Dominicis function of the form [9].

G*(t) is the overlap at time t between two wave functions of the hidden Fermi liquid: one in which the site i was emptied of down-spin electrons at time t=0, and the electrons thereafter propagated according to the free-particle Hamiltonian; and the other in which the site was only emptied at time t.

I argue that the first wave function is equivalent to that which would result from turning on a local potential at t=0 which would remove the correct number of down-spin electrons from the site, that is one which, using the Friedel theorem [8], would establish a phase shift η giving us the proper δN. With constant (Fermi) velocity propagation, the resulting phase displacement of the electron waves travels outward as a more or less spherical shell, with the phase of the waves inside of the shell displaced relative to those outside. The overlap of such a wave function with the unperturbed ground state is precisely the NDD expression. The number of down-spin electrons below the Fermi level, per site, is δN=(1-x)/2, so that according to [9] the power law should be p=(1-x)$^2$/8.

This is, however, not quite right. When we examine the "hat" operators of equation 11, we note that they can be factorized in two different manners:

$$\hat{c}_{i\sigma} = c_{i\sigma} c_{i-\sigma} c^*_{i-\sigma} = c_{i-\sigma} S_i^- \text{ or } = c_{i\sigma}(1 - n_{i-\sigma}) \quad [13]$$

When we write these two factorizations out in momentum space they look rather different.

$$\hat{c}_{k,\sigma} = \sum_j e^{ir_j \cdot k} \hat{c}_{j,\sigma} = \sum_q c_{k-q,-\sigma} \sum_{k'} c^*_{k'-q,-\sigma} c_{k',\sigma}$$

$$or = \sum_q c_{k-q,\sigma} \sum_{k'} c_{k',-\sigma} c^*_{k'-q,-\sigma} \quad [14]$$

We use two observations about the effect of the exclusion principle and conservation of energy on electron-electron scattering for states close to a Fermi surface. One is that as we squeeze down to the Fermi surface only forward scattering survives (see fig 6.1 of reference 6) It is this forward scattering, for instance, which causes the mean-field interactions in Fermi liquid theory. The second observation is that inelastic scattering only conserves energy when the Fermi velocities match (since the energy difference is $(v_F - v_F').q$ ) . Thus inelastic scatterings can be roughly described as generating soft tomographic Tomonaga bosons, with the incoming particle losing momentum q and generating a Tomonaga boson of that momentum, and the two terms of [14] create two different sets of bosons, carrying density and spin density respectively. These are two separate scattering channels with the same amplitude in each, and according to the rule of equation [9], they will add, doubling the exponent p. It is this doubled value which we have used in fitting to experimental data[10].

For comparison with ARPES and transport data it is necessary to derive the momentum-dependent Green's function. As pointed out in reference 7, given the Green's function for the symmetric channel—spherical waves, in the ideal case—it is trivial to derive the plane wave result, using the ancient device of Huygens' construction. The Green's function for plane wave propagation is simply the linear superposition of spherical waves emanating from each point on the previous plane wave front. Thus if the correction for the incoherent radiation of Tomonagons is the simple function of time, constant×$t^{-2p}$, this factor will be common to every component and will appear in the plane wave Green's function, multiplied by the free quasiparticle Green's function.

Motion of a particle near the Fermi surface is essentially one-dimensional, so we may take the free particle Green's function in space-time as $1/(x-v_F t)$. In order to get the imaginary part of G (the

density) in k and frequency space, we must Fourier transform G(x,t):

$$G(k,\omega) = \int dxdt\, e^{i(kx-\omega t)} t^{-p}/(x - v_F t)$$

doing the x integration by a contour integration,

(the sign of t determines which way to close the contour)

this becomes $\int t^{-p} dt\, e^{i(v_F k - \omega)t} \propto ([v_F k - \omega])^{-1+p}$

[15]

The imaginary part of this expression is the T=0 EDC. If p=0, this is just a delta function at the quasiparticle energy; but if p is finite it has an imaginary part for all $\omega > v_F k$ : the quasiparticle becomes a cut singularity, not a pole, in the complex plane, and does not have a finite residue *at* the point of singularity, i e has Z=0.

Understanding the thermal and transport behavior involves thought about the physical meaning of this situation. These properties were discussed at some length in reference 6, and I believe that although the assumption of spin-charge separation which underlay that discussion was too radical, the physical consequences do not differ very much from what was claimed there.

In conventional Fermi liquid theory, the spectral properties do not have much to do with the thermal properties, even though β=1/T plays the formal role of an imaginary time in many-body perturbation theory.[11] But one of the striking things about the strange metal phase is that the conductivity as well as the Green's functions as seen in ARPES obey "ω, T scaling" where the T and frequency dependences seem to scale together.

There are actually separate arguments for the two types of response functions. In FLT the Green's functions have simple poles at the quasiparticle energies, which are in principle at sharp energies; it is only the existence of scattering which gives the quasiparticle a breadth and moves the pole off the real axis. The Green's functions as complex functions in spacetime can be defined, at finite T, to be (anti)periodic in the imaginary time direction with period 1/T, and the resulting array of poles along the imaginary axis turn out to be simply the poles of the Fermi function as a function of complex energy $\omega$, and do not imply any decay or scattering.

$$f(\omega/T) = (1 + e^{-\omega/T})^{-1} = 1 + \tanh(\omega/2T)$$

which is a periodic functon of $i\omega$ with poles at $\omega_n = i(2n+1)T$. Yuval showed many years ago[12] that the Nozieres-deDominicis function indeed has a periodic structure in imaginary time:

$$t^{-p} \to [\sinh(\pi T t)/\pi T]^{-p} \quad [16]$$

It makes the manipulations very much easier to approximate this expression (quite accurately) simply by

$$G*(t) \cong t^{-p} e^{-\Gamma t}, with\, \Gamma = AT, A \cong \pi p \quad [17]$$

We see that the well-known "linear T" decay follows from the power law and the analytic structure of Green's functions. Note that $\Gamma$ vanishes as $p \to 0$.

Using the approximation [17] it becomes quite easy to Fourier transform the Green's function and to obtain an expression for the ARPES intensity:

$$EDC = \text{Im}\{G\} \propto \text{Im}\{\frac{f(\omega/T)}{[(v_F k - \omega) + i\Gamma]^{1-p}}\} =$$

$$= f(\omega/T) \frac{\sin(1-p)\cot^{-1}([\omega - v_F k]/\Gamma)}{[(v_F k - \omega)^2 + (\Gamma)^2]^{(1-p)/2}}$$

[18]

(The Fermi distribution is natural since all final states are single-Fermion. It comes from the space Fourier transform of $G_0$.) This spectrum is not unfamiliar to x-ray spectroscopists: except for the Fermi distribution factor, it is the Doniach-Sunjic[13] line shape of x-ray line spectra in metals.

[18], except for one further adjustment, is the expression which has been compared with the laser ARPES results of Koralek and Dessau[14] in a recent publication. We have no reason to believe that the underlying, hidden Fermi liquid does not have the conventional electron-electron umklapp scattering. For the standard exclusion principle reasons this vanishes at the Fermi surface but will give us an addition to $\Gamma$ proportional to $(k-k_F)^2$. In fact, the measured $\Gamma$ fit that prescription remarkably well –see Fig 1, borrowed from reference 13.

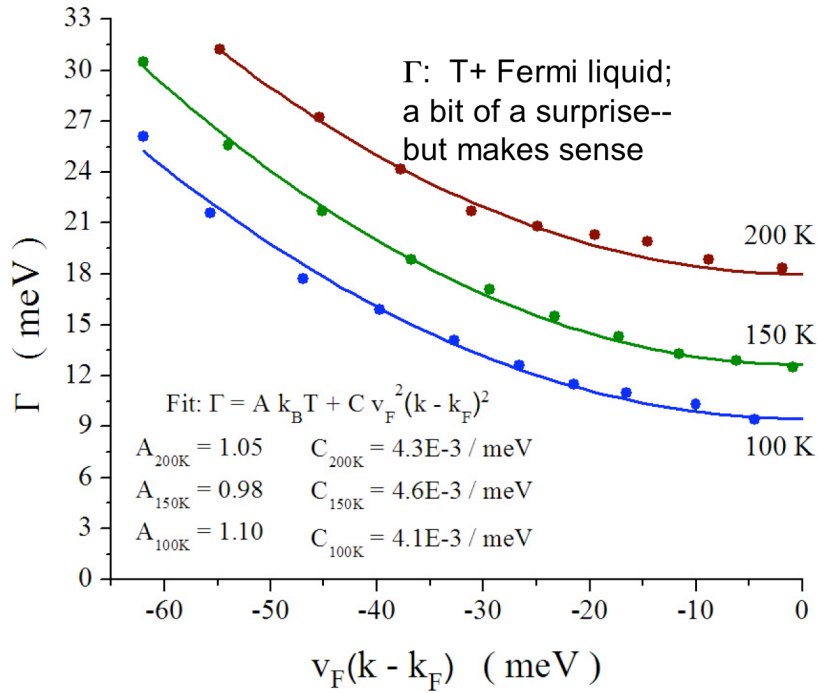

**Figure1:** Γ values from fits of equation 18 as a function of momentum and temperature. Curves are fits to the sinple expression in the insert.

The experimental infrared conductivity in the strange metal region fits remarkably well to a simple power law dependence on frequency over a wide range from T≅10 mev to at least 300 mev. This was discovered by Schlesinger and Collins[15] and has been measured with great accuracy by van der Marel and others. There is a corresponding ω,T scaling "linear T" resistivity which was the first manifestation of the strange metal which was noticed. The interpretation of these results which is favored is that in this regime the conductivity is dominated by the lowest-order "bubble" diagram involving the simple product of two G's, one each for the electron and the hole excited by a long-wavelegth photon. For this to be the case "vertex corrections" must be negligible, in spite of the fact that the processes we have been discussing are primarily forward electron-electron scattering and, unaided, cannot cause any resistivity at all. This situation, for the resistivity, was discussed in a paper by Ogata and myself[16] and is resolved by the concept of "drag" and "non-drag" regimes. For the infrared conductivity there is no problem: the hole and electron survive so briefly they have no chance to interact—equivalently, lowest-order perturbation theory suffices. But for the resistivity, the electron decay process will be ineffective if the current can be carried by the "hidden" Fermi liquid excitations; the relevant scattering rate will then be that of this Fermi liquid. I believe that it is this crossover, rather than a crossover to a true Fermi liquid, which occurs as we move into the overdoped regime.

HIDDEN FERMI LIQUID IN THE SUPERCONDUCTING REGIME.

As we have already noted, in the superconductor the $t^{-p}$ decay of the Green's function is cut off at times comparable to the inverse energy gap and there is a coherent, if attenuated, quasiparticle. The Green's function and ARPES spectrum can be calculated by methods used to treat the x-ray spectra of superconductors by Yongjun Ma.[17] (I am indebted to V Muthukumar for calling my attention to this paper.) P Casey and the author have applied these methods and produced tentative spectra; unfortunately these do not resemble the experimental data very closely, apparently because of the great inhomogeneity of the energy gaps in BISSCO. This work will be the subject of a later paper.

ACKNOWLEDGEMENTS